\documentclass[preprint,showpacs,preprintnumbers,amsmath,amssymb]{revtex4}
\usepackage{graphics}
\usepackage{dcolumn}
\usepackage{bm}

\begin{document}

\title{Physical mechanism of superluminal traversal time: interference between multiple finite wave packets}
\author{Xi Chen$^{1}$\footnote{Email address: xchen@shu.edu.cn} and Chun-Fang Li$^{1,2}$\footnote{Corresponding author. Email address: cfli@shu.edu.cn}}

\affiliation{$^1$Department of Physics, Shanghai University,
Shanghai 200444, People's Republic of China} \affiliation{$^2$State
Key Laboratory of Transient Optics Technology, Xi'an Institute
of\\Optics and Precision Mechanics, Academia Sinica, Xi'an 710119,
People's Republic of China}

\date{\today}

\begin{abstract}
The mechanism of superluminal traversal time through a potential
well or potential barrier is investigated from the viewpoint of
interference between multiple finite wave packets, due to the
multiple reflections inside the well or barrier. In the case of
potential-well traveling that is classically allowed, each of the
successively transmitted constituents is delayed by a subluminal
time. When the thickness of the well is much smaller in comparision
with a characteristic length of the incident wave packet, the
reshaped wave packet in transmission maintains the profile of the
incident wave packet. In the case of potential-barrier tunneling
that is classically forbidden, though each of the successively
transmitted constituents is delayed by a time that is independent of
the barrier thickness, the interference between multiple transmitted
constituents explains the barrier-thickness dependence of the
traversal time for thin barriers and its barrier-thickness
independence for thick barriers. This manifests the nature of
Hartman effect.
\end{abstract}

\maketitle

The question of how long it takes for quantum particles to tunnel
through a potential barrier has attracted considerable attention for
many decades \cite{Hauge-S,Chiao-S,Muga-SE,Nimtz}. Theoretical
investigations
\cite{Hartman,Buttiker-L,Martin-L,Steinberg-Chiao,Martinez} and
experimental researches \cite{Enders-1,Steinberg-Kwiat,Sp-R,Balcou}
have shown that the traversal time defined by group delay, also
known as the phase time in the literature \cite{Hauge-S}, which
describes the motion of a wave packet peak \cite{Wigner}, has the
well-known superluminality. Though the group delay for quantum
particles tunneling through a potential barrier depends on the
barrier thickness for thin barriers, it is saturated to a constant
for thick barriers. This is known as the ``Hartman effect"
\cite{Hartman,Martinez}. It was further shown that quantum particles
traveling through a quantum well can be advanced, so that the group
delay in transmission can be negative \cite{Li-Wang,Muga,Chen-Li}.
This counterintuitive phenomenon has been experimentally
demonstrated in mirocwave analogy experiment \cite{Vetter}. In
addition, superluminal propagation was also theoretically predicted
\cite{Garrett} and experimentally verified \cite{Chu,Wang-1} for
light pulses through anomalous dispersion media.

Nevertheless, the mechanism for superluminal traversal time remains
controversial to the present day \cite{Buttiker-1,Winful}. Wang and
Zhang \cite{Wang-Z} attributed the superluminal traveling in the SKC
experiment \cite{Steinberg-Kwiat} to the reshaping due to the
interference between uncertain path modes. Japha and Kurizki
\cite{Japha} demonstrated, by use of the impulse response function,
that the destructive interference between accessible causal paths
plays a key role in superluminal time delays. Others
\cite{Dogariu,Wang-LLZ,Guo} regarded it as the reshaping owing to
the interference between different frequency components that undergo
different phase shifts. Recently, Winful \cite{Winful-1} explained
the Hartman effect on the basis of saturation of the integrated
probability density under the barrier. More recently, Martinez
\cite{Martinez} investigated the origin of Hartman effect by the
interference between the evanescent waves in the region of barrier.

In this Letter, we give a clear picture of the reshaping mechanism
of the transmitted wave packet to explain the superluminal traversal
time. The reshaping of the whole transmitted wave packet results
from the coherent interference between its different constituents,
which are not plane wave components but all finite temporal wave
packets, and arise from multiple reflections inside the well or
barrier. In the case of potential-well traveling that is classically
allowed, each of the finite constituents is temporally delayed by a
subluminal time. We also show that the reshaping does not change the
profile of the wave packet as long as a condition is satisfied. On
the other hand, in the case of potential-barrier tunneling that is
classically forbidden, each of the successively transmitted
constituents is delayed by a time that is independent of the barrier
thickness.

Let us first consider the traversal time of quantum particles
through a one-dimensional rectangular potential well of depth
$V_{0}$, extending from $0$ to $a$. A temporal wave packet can be
expressed in terms of plane wave components as $\Psi_{i}(x,t)=
\int_{-\infty}^{+\infty} A(\omega) \exp[i(kx-\omega t)] d\omega$,
where $\omega=E/\hbar$, $k=(2 \mu E)^{1/2}/\hbar$, $\mu$ is the mass
of the particle, and $E$ is the energy. For a Gaussian-shaped
incident wave packet whose peak is assumed to locate at $x=0$ and
$t=0$,
\begin{equation} \label{incident beam}
\Psi_{i}(0,t)=\exp{\left(-\frac{t^2}{2\tau^2}\right)}\exp{(-i \omega_0 t)},
\end{equation}
its amplitude spectral distribution is also a Gaussian function,
$A(\omega)=\tau \exp[-(\tau^2/2)(\omega-\omega_{0})^2]$, around a
central energy $E_0=\hbar \omega_0$, where $\tau$ is  half the width
of the wave packet. For a frequency component $\exp[i(kx-\omega
t)]$, letting be $T(\omega) \exp\{i[k(x-a)-\omega t]\}$ the
corresponding transmitted component, the amplitude transmission
coefficient $T=e^{i\Phi}/f$ is determined, according to
Schr\"{o}dinger equation and boundary conditions, by the following
complex number \cite{Li-Wang},
$$
f e^{i \Phi}= \cos k' a+ \frac{i}{2}\left(\frac{k}{k'}+ \frac{k'}{k}\right)\sin k' a,
$$
where $1/f$ is the absolute value of $T$, and $k'= [2 \mu
(E+V_{0})]^{1/2}/\hbar$. Obviously, the phase angle, $\Phi$, of the
transmission coefficient is the phase shift of the transmitted wave
at $x=a$ with respect to the incident wave at $x=0$, resulting
directly from the multiple transmissions due to the multiple
reflections inside the potential well. As a matter of fact, the
phase shift in transmission would be $k' a$ if only the first
transmitted wave were considered. Taking into account multiple
reflections and transmissions at the well sides, the transmission
coefficient can be rewritten as the series expansion,
$$
T(\omega)=\sum\limits_{j=1}^\infty T_{j} (\omega),
$$
with
$$
T_{1} (\omega)= \frac{4kk'}{(k'+k)^2}\exp{(i k'a)},
$$
\begin{equation} \label{Tj}
T_{j} (\omega)= T_{1}\left[\left(\frac{k'-k}{k'+k}\right)\exp(i k'a)\right]^{2(j-1)}.
\end{equation}
This exponential series is the same as is encountered in multiple
transmissions through the Fabry-Perot interferometer \cite{Born}.
Therefore, the transmitted wave packet at $x=a$
\begin{equation} \label{transmission}
\Psi_t (a,t)= \int_{-\infty}^{+\infty} T(\omega) A(\omega)\exp (-i \omega t) d\omega
\end{equation}
can be represented by
\begin{equation}
\label{full sum} \Psi_t (a,t) = \sum^{\infty}_{j=1} \Psi_{j}(a,t),
\end{equation}
where
\begin{equation} \label{mulitple transmitted WP}
\Psi_{j}(a,t)= \int_{-\infty}^{+\infty} T_{j} (\omega) A(\omega) \exp (-i \omega t) d\omega.
\end{equation}
Assuming $|T_j (\omega)|$ to be a constant, $|T_{j0}|=|T_j
(\omega_0)|$, in the interval in which $A(\omega)$ is appreciable
and employing the first-order Taylor expansion of $k'$ at
$\omega=\omega_0$, $ k'\approx k'_0 - (\partial k'/\partial
\omega_0)(\omega-\omega_0)$, where $\partial/\partial \omega_0$
denotes the derivative with respect to $\omega$ evaluated at
$\omega_0$, we get for the wave packet (\ref{mulitple transmitted
WP}) at $x=a$,
\begin{equation} \label{partial Gaussian wavepacket}
\Psi_{j} (a,t) \approx |T_{j0}|
\exp{\left[-\frac{(t-t_{j})^2}{2\tau^2}\right]}\exp{(-i \omega_0
t)},
\end{equation}
where $t_{j}=(2j-1)t_1$, $t_1=a/v_g $, and $v_g=\hbar k'_0/\mu$ is the group velocity in the
region of the potential well. It is noted that the subscript $0$ in this paper denotes values
taken at $\omega=\omega_0$. Comparison with Eq. (\ref{incident beam}) shows that $\Psi_{j} (a,t)$
has the same shape as that of the incident wave packet and is temporally delayed by a subluminal
positive time, $t_j$, due to the fact that the traveling through a potential well is classically
allowed.

\begin{figure}[]
\scalebox{0.34}[0.32]{\includegraphics{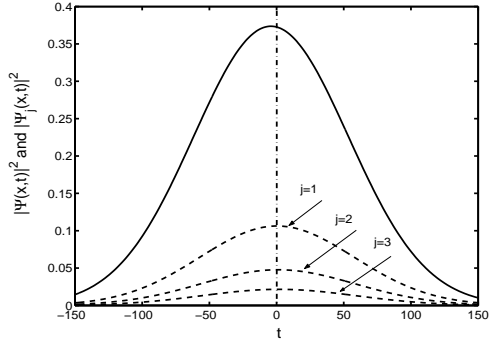}}
\caption{\label{comparison1} Temporal relation of transmitted wave
packet (solid curve) with its first three constituents (dotted
curves as marked) at $x=a$, where $E_0=0.01V_0$, $a=3.4/k'_0$, and
the time axis is in units of $\tau_u=\mu/\hbar k_0 k'_0$. The half
width of the incident wave packet is chosen to be $\tau=80 \tau_u$.}
\end{figure}

\begin{figure}[]
\scalebox{0.34}[0.32]{\includegraphics{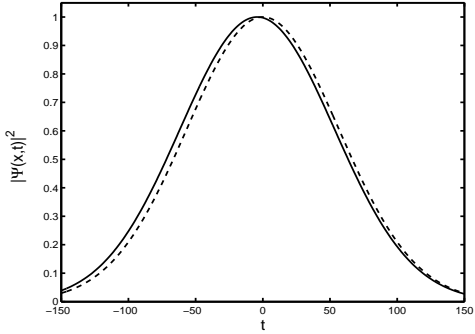}}
\caption{\label{comparison2} Comparison between the normalized
profile of transmitted wave packet at $x=a$ (solid curve) with that
of incident wave packet at $x=0$ (dashed curve), where all the
physical parameters are the same as in Fig. \ref{comparison1}.}
\end{figure}

According to Eqs. (\ref{full sum}) and (\ref{partial Gaussian
wavepacket}), the whole transmitted wave packet is the coherent
superposition of its successively transmitted constituents with
subluminal delay time, each of which is delayed from its adjacent
one by a positive time $2t_1$. However, numerical calculations have
shown that if $2t_1$ is much smaller than $2 \tau$, the time width
of the incident wave packet, that is to say, if the condition
\begin{equation}\label{restriction-a}
a \ll v_{g} \tau
\end{equation}
holds for the incident wave packet, the whole transmitted wave
packet will maintain the shape of incident wave packet with
superluminal traversal time. In Fig. \ref{comparison1} is depicted
the temporal relation of the whole transmitted wave packet (solid
curve) with its first three constituents (dotted curves as marked)
at $x=a$. Since the half width of the incident wave packet is chosen
to be $\tau=80 \tau_u$, we have $a/(v_g \tau)= k_0 a/80 \approx
0.0425$, which satisfies the condition (\ref{restriction-a}). It is
clearly shown that the traversal time of the whole transmitted wave
packet is negative though the delay time of the successively
transmitted constituents are positive and equal to $(2j-1) t_1$.
This tells us a fact that the interference between wave packets of
subluminal time delay may produce superluminal time delay. Fig.
\ref{comparison2} further shows negligible distortion of the
transmitted wave packet at $x=a$ (solid curve) in comparison with
the incident wave packet at $x=0$ (dashed curve), where all the
physical parameters are the same as in Fig. \ref{comparison1}.

\begin{figure}[]
\scalebox{0.34}[0.32]{\includegraphics{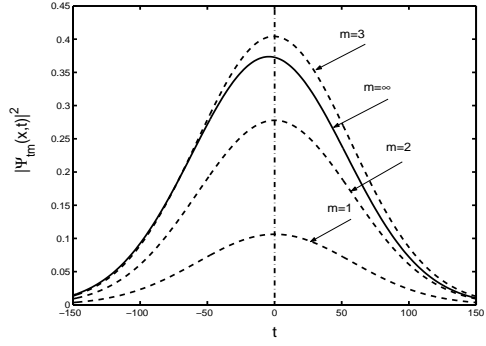}}
\caption{\label{reshaping process} Temporal relation between the
wave packets that are defined by Eq. (\ref{partial sum}), where the
solid curve corresponds to the whole transmitted wave packet
($m=\infty$), and the dashed curves correspond to $m=1,2,3$ as
marked. The physical parameters are the same as in Fig.
\ref{comparison1}.}
\end{figure}

In order to have a clear look at the effect of the interference
between successively transmitted constituents on the superluminal
time delay, we may define a wave packet at $x=a$ as follows,
\begin{equation} \label{partial sum}
\Psi_{tm} (a,t)= \sum^{m}_{j=1} \Psi_{j}(a,t).
\end{equation}
When $m \rightarrow \infty$, it gives us the whole transmitted wave
packet (\ref{full sum}). In Fig. \ref{reshaping process} is shown
the temporal relation of so defined wave packets for $m=1,2,3$ and
$m=\infty$. It reveals how the whole transmitted wave packet is
gradually reshaped and how the superluminal and even negative
traversal time is gradually established by the interference between
its successively transmitted constituents that are all temporally
delayed by a subluminal time. This reshaping is related to
Aharonov's weak measurements of the time delay
\cite{Aharonov,Sokolovski}.

In fact, when the condition (\ref{restriction-a}) is satisfied, the
absolute value and phase of transmission coefficient, $T(\omega)$,
is linearly dependent on $\omega$ within the interval in which the
spectral distribution, $A(\omega)$, is appreciable. It is thus
reasonable to express $T(\omega)$ as an exponential form, expand the
exponent in Taylor series at $\omega_0$, and retain up to the
first-order term to obtain
\begin{equation} \label{first-order expansion}
T(\omega)= \exp [\ln T(\omega)] \approx T_0 \exp
           \left[
                  \frac{1}{T_0} \frac{dT}{d \omega_0}(\omega-\omega_0)
           \right].
\end{equation}
Introducing two real parameters $\tau_n$ and $\tau_{\phi}$ defined
by
\begin{displaymath}
\tau_n+ i \tau_{\phi}= \frac{1}{T_0} \frac{d T}{d\omega_0},
\end{displaymath}
and substituting Eq. (\ref{first-order expansion}) into Eq. (\ref{transmission}), we get for the
transmitted wave packet at $x=a$,
\begin{equation} \label{field of transmitted wave packet}
\Psi_{t} (a,t)= C \exp \left[ -\frac{(t-\tau_{\phi})^2}{2 \tau^2} \right]
                  \exp \left[-i(\omega_0+\frac{\tau_n}{\tau^2}) t \right],
\end{equation}
where
$$
C=\sqrt{2 \pi} T_0 \exp \left( \frac{\tau^2_n}{2 \tau^2} \right)
                   \exp \left(i\frac{\tau_n \tau_{\phi}}{\tau^2} \right).
$$
The first exponential factor on the right-handed side of Eq.
(\ref{field of transmitted wave packet}) shows that the transmitted
wave packet has the same shape as that of incident wave packet with
traversal time $\tau_{\phi}=\partial \Phi/\partial \omega_0$, which
can be negative as well as positive \cite{Li-Wang,Chen-Li,Muga}. The
second exponential factor indicates that the central energy of the
transmitted wave packet moves to a different value \cite{Aquino},
$\hbar (\omega_0+ \tau_n/\tau^2)$, other than $\hbar \omega_0$.

On the contrary, when the condition (\ref{restriction-a}) is not
satisfied, numerical calculations have shown that the transmitted
wave packet is distorted significantly, due to the fact that the
absolute value and phase of transmission coefficient do not linearly
depend on $\omega$ in the interval in which the spectral
distribution, $A(\omega)$, is appreciable. Indeed, the concept of
group delay in this case is no longer applicable.

Next, let us look at the traversal time in barrier-tunneling
process. For a potential barrier of height $V_0$ and thickness $a$
and an incident wave packet whose central energy $E_0$ is smaller
than $V_0$, the above discussed properties of transmission
coefficient is valid when $k'$ is replaced by $i \kappa$, where
$\kappa= [2 \mu (V_0-E)]^{1/2}/\hbar$. In this case, Eq.
(\ref{mulitple transmitted WP}) represents the successively
transmitted constituents due to multiple reflections of evanescent
wave \cite{Martinez} inside the region of potential barrier.
Introducing two positive parameters, $\Delta$ and $\delta$, defined
by $k+i \kappa= \Delta \exp(i \delta)$, we have
$$
T_1=\frac{4k \kappa}{\Delta^2} \exp(-\kappa a) \exp(i \frac{\pi}{2}) \exp(-2i \delta),
$$
and
$$
T_j=[\exp(- \kappa a) \exp(-2i \delta)]^{2(j-1)}T_1.
$$
It is clear that the phase of each of transmission coefficients,
$T_j$, is independent of barrier thickness. As a result, each of the
wave packet constituents, $\Psi_j$, is delayed from the incident
wave packet by a time that is independent of the barrier thickness.
It is also shown that the magnitude of each of $T_j$ is decayed from
that of its preceding counterpart by a factor $\exp(2 \kappa a)$.

When the barrier is thin in comparison with the penetration depth,
$1/\kappa$, the time delay of the whole transmitted wave packet
depends on the barrier thickness. This dependence is rooted in the
interference between the multiple transmitted constituents. When the
barrier is thick enough, $\kappa a \gg 1$, each of transmitted
constituents is very small and can be neglected in comparison with
its preceding one, although the interference between them still
exists. As a result, the whole transmitted wave packet in this case
is approximately equal to the first transmitted constituent,
$$
\Psi_{t}(a,t) \approx \Psi_{1}(a,t)= \int_{-\infty}^{+\infty} T_{1} (\omega) A(\omega) \exp (-i
\omega t) d\omega.
$$
So the traversal time is equal to the time delay of the first
transmitted constituent, and is given by, in terms of the phase of
$T_1$,
$$
\tau_{\phi}=\frac{d(-2 \delta+ \pi/2)}{d \omega_0}=\frac{2 \mu}{\hbar k_0 \kappa_0}.
$$
It is obviously independent of the barrier thickness. This
conclusion results from the fact that no phase accumulates in the
barrier region, and the phase shift only comes from the boundary of
barrier \cite{Enders-1,Aharonov}. All those amount to manifest the
nature of Hartman effect in barrier tunneling.

In summary, the superluminal traversal time of a quantum wave packet
through a potential well has been explained by the mechanism of
reshaping due to the interference between its successively
transmitted wave packet constituents that arise from the multiple
reflections inside the potential well. A physical condition was
advanced that is required for the transmitted wave packet to
maintain the shape of incident wave packet. This mechanism was
further used to show the nature of Hartman effect in barrier
tunneling. We believe that the reshaping mechanism of wave packets
discussed here is also applicable to other phenomena, such as the
Goos-H\"{a}nchen displacement in frustrated total internal
reflection \cite{Steinberg-Chiao, Balcou} and the anomalous lateral
displacement \cite{Li-Wang} of light beam transmitted through a thin
slab of optically thick medium.

\section*{Acknowledgments}
This work was supported in part by the National Natural Science
Foundation of China (Grants 60377025 and 60407007), Science and
Technology Commission of Shanghai Municipal (Grants 03QMH1405 and
04JC14036), and the Shanghai Leading Academic Discipline Program
(T0104).

\end{document}